\pdfoutput=1
\documentclass[conference, a4paper, 9pt, twoside]{IEEEtran}
\usepackage[english]{babel}
\usepackage[T1]{url}
\usepackage[final]{microtype}
\usepackage{amsmath}
\usepackage[romanConstants, boldVectors]{scstuff}
\usepackage{units}
\usepackage{acronym}
\usepackage{cite}
\usepackage[caption=false, subrefformat=parens]{subfig}
\usepackage[pdftex]{graphicx}
\usepackage{booktabs}
\usepackage[inline]{enumitem}
\usepackage[pdftex, unicode, bookmarks=false, bookmarksnumbered=true,
  pdfpagemode={UseOutlines}, plainpages=false, pdfpagelabels=true,
  colorlinks=true, linkcolor={DarkBlue}, citecolor={DarkBlue}, 
  urlcolor={DarkBlue}]{hyperref}
\usepackage{ucs}
\usepackage[utf8x]{inputenc}
\usepackage[LGR,T1]{autofe}
\usepackage{textcomp}
\usepackage{mdwtab}
\usepackage[svgnames]{xcolor}
\usepackage[textsize=footnotesize, textwidth=1cm]{todonotes}
\usepackage{tikz}
\usetikzlibrary{arrows}

\renewcommand\normalsize{\fontsize{8.8}{10.7}\selectfont}

\graphicspath{{./Figures/}}

\makeatletter
\let\thanks\@IEEESAVECMDthanks
\makeatother

\makeatletter
\newcommand\ifacroused[3]{%
  \expandafter\ifx\csname ac@#1\endcsname\AC@used
  #2\else #3\fi}
\makeatother
\newacro{RNG}{Random Number Generator}
\newacro{USB}{Universal Serial Bus}
\newacro{I2C}[I\textsuperscript{2}C]{Inter Integrated Circuit}
\newacro{SPI}{Serial Peripheral Interface}
\newacro{USART}{Universal Synchronous Asynchronous Receiver Transmitter}
\newacro{ADC}{Analog to Digital Converter}
\newacro{DAC}{Digital to Analog Converter}
\newacro{uC}[μC]{microcontroller}
\newacro{OS}{Operating System}
\newacro{PWAM}{Piece-Wise Affine Markov}
\newacro{IoT}{Internet of Things}
\newacro{TH}[T/H]{Track-and-Hold}
\newacro{OA}{Operational Amplifier}
\newacro{lsb}{least significant bit}
\newacro{msb}{most significant bit}

\newcommand{\mailto}[1]{\href{mailto:#1}{\nolinkurl{#1}}}

\DeclareMathOperator{\ADC}{ADC}
\DeclareMathOperator{\DAC}{DAC}

\title{Very Low Cost Entropy Source Based on Chaotic Dynamics Retrofittable on
  Networked Devices to Prevent RNG Attacks}%
\author{%
  \IEEEauthorblockN{Mattia Fabbri}
  \IEEEauthorblockA{University of Bologna, Italy\\
    \mailto{mattia.fabbri2@studio.unibo.it}}%
  \and \IEEEauthorblockN{Sergio Callegari}
  \IEEEauthorblockA{ARCES/DEI, University of Bologna, Italy\\
    \mailto{sergio.callegari@unibo.it}}%
  \thanks{This is a post-print version of a paper published in the
    Proceedings of 2014 IEEE International Conference on Electronics, Circuits,
    and Systems (ICECS). 
    Cite
    as:\protect\\[1ex]
    Fabbri M., Callegari S., ``Very Low Cost Entropy Source Based on Chaotic
    Dynamics Retrofittable on Networked Devices to Prevent RNG Attacks,''
    \emph{IEEE 21th International Conference on Electronics, Circuits, and
      Systems (ICECS 2014),} pp.~175--178, Dec.\@ 2014.
    \protect\\[1ex]
    Copyright © 2014 IEEE. Personal use of this material is permitted. However,
    permission to use this material for any other purposes must be obtained
    from the IEEE by sending a request to
    \mailto{pubs-permissions@ieee.org}.
    \protect\\[-2ex]}}

\initializeplaintitle
\initializeplainauthor[%
\def\and{, }
\def\IEEEauthorblockN#1{#1\ignorespaces}%
\def\IEEEauthorblockA#1{}]
\hypersetup{pdftitle=\plaintitle, 
  pdfauthor=\plainauthor,
  pdfsubject={},
  pdfkeywords={}}

\begin{document}
\maketitle
\bstctlcite{default:BSTcontrol}

\begin{abstract}
  Good quality entropy sources are indispensable in most modern cryptographic
  protocols. Unfortunately, many currently deployed networked devices do not
  include them and may be vulnerable to \ac{RNG} attacks. Since most of these
  systems allow firmware upgrades and have serial communication facilities, the
  potential for retrofitting them with secure hardware-based entropy sources
  exists. To this aim, very low-cost, robust, easy to deploy solutions are
  required. Here, a retrofittable, sub 10\$ entropy source based on chaotic
  dynamics is illustrated, capable of a \unit[32]{kbit/s} rate or more and
  offering multiple serial communication options including \acs{USB},
  \acs{I2C}, \acs{SPI} or \acs{USART}. Operation is based on a loop built
  around the \ac{ADC} hosted on a standard microcontroller.
\end{abstract}
\acresetall

\section{Introduction}
One of the most powerful trends recently seen in technology is the networking
of a large number of devices originally operating in isolation. By 2020 more
than 25 billion devices will be interconnected \cite{Gartner:IOT-2013}. This
\ac{IoT} can bring smartness to many environments whose items gain the ability
to cooperate and a richer set of information to base their behaviors upon. At
the same time, making a huge number of devices remotely reachable poses
profound security challenges because of the immense attack surface
\cite{roman:computer-44-9}.

Unfortunately, many devices being brought online lack certain security-oriented
subsystems. This is due to cost and conception, since networking is often
applied to mundane items and awareness on security issues is still growing. In
a short time, there will be a huge legacy of networked devices missing hardware
parts critical to security. This can be long lasting since humble devices lack
any upgrade lust and are only substituted when they break, as users often fail
to see potential threats as lack of functionality. Surveys reveal that about
half of the networking devices found in businesses are already on the brink of
obsolescence \cite{Lynn:PCMAG-2011-05}. To contrast this phenomena, some
authors suggest that embedded networking units should be given a scheduled
end-of-life \cite{Geer:SOT-2014}. Until this is a reality (if ever), the
alternative is to minimize the costs of patching current devices, either in
manufacturing, through design revisions, or on field, through software upgrades
and hardware retrofits.

\begin{figure}
  \begin{tightcenter}
    \includegraphics[width=0.5\lw]{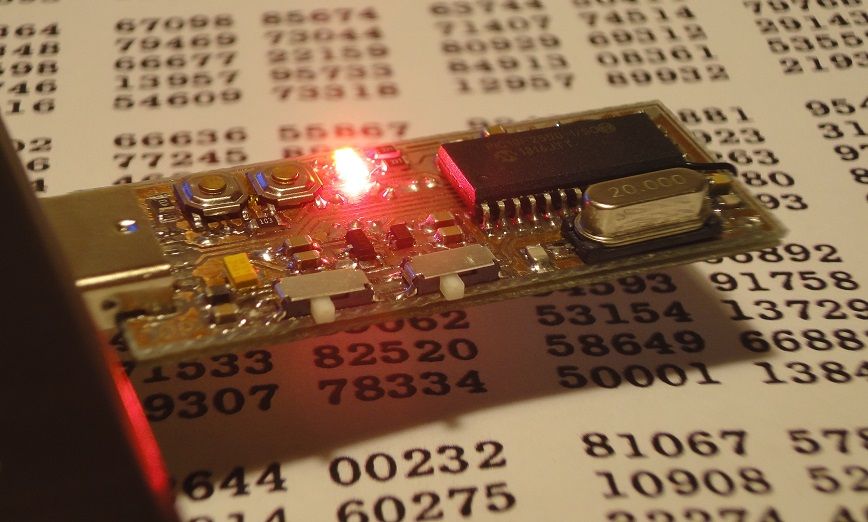}
  \end{tightcenter}
  \caption{Prototype of self-contained entropy source with \ac{USB} port,
    including diagnostic LEDs, control switches, trimmers, test-pads, and a DIP
    \ac{uC} package significantly enlarging the layout.}
  \label{fig:demo}
\end{figure}

This paper illustrates the design of a hardware primitive critical for
security, namely an entropy source, whose absence can favor \ac{RNG} attacks
(references to past incidents in \cite{wikipedia:RNGattack}). True-\acp{RNG}
and hardware entropy sources are becoming mainstream in computers
\cite{taylor:spectrum-48-9}, but are still missing from most low-cost/embedded
devices. The current proposal can be declined in multiple forms: from a
self-contained plug-in applicable on field to existing systems
(Fig.~\ref{fig:demo}) to a schematic upgrade re-using an existent \ac{uC},
targeting manufacturers. In either case, key features are: (i) very low cost
(sub-10\$ for a self-contained unit with \ac{USB} connector, sub-5\$ for
designs where a \ac{uC} is available to share); and (ii) high interconnection
flexibility (\ac{USB}, \acs{I2C}, \acs{SPI} or \acs{USART} interfaces,
depending on the adopted \ac{uC}) in order to favor adoption as a retrofit. The
entropy rate that can be delivered is in the few tens \unit{kbit/s} range,
depending on the \ac{uC} (over \unit[32]{kbit/s} have been experimentally
observed with a PIC18F2550).

Similar solutions are scarce on the market, with true-\ac{RNG} retrofits
ranging from near 40\$ to over 1000\$ (i.e., at far higher costs, even
considering selling margins). Closest competitors are the TRNG98 modules
\cite{DomstedtElectronics:TRNG9880}, and the Simtec Entropy Key. Differently
from any other design, the current proposal is based on chaotic dynamics,
rather than direct amplification of electronic noise (thermal, avalanche,
etc.), which can make the system less susceptible to tampering and side channel
attacks \cite{Pareschi:ISCAS09}.  It is not the first time that chaos is
proposed for \acp{RNG} \cite{Callegari:IJCTA-33-1}, but the current design has
the distinguished feature of re-using \acp{ADC} readily available on
\acp{uC}. \ac{ADC} reuse, originally introduced in \cite{Callegari:TSP-53-2},
has already seen experimental validation \cite{Pareschi:ISCAS09}, but so far
only on full-custom integrated circuits with ad-hoc designed \emph{pipeline}
\acp{ADC}, where stages can be separated from each other.  Here, taking
advantage of the generalized theory in \cite{Callegari:ISCAS-2007,
  Callegari:ISCAS-2008}, the approach is used for the first time on an
unmodified commercial \ac{ADC}, with a successive approximation architecture
and harsh operating conditions (\ac{uC} \acp{ADC} are often poorly isolated
from digital noise). This setup requires some specific design arrangements that
are here thoroughly detailed.

\section{Background: \acp{RNG} and security}
\label{sec:background}
Cryptographic systems base their operation on the existence of some secret data
known to authorized users and unpredictable by others. To warrant
unpredictability, random strings are often employed, e.g., in \emph{keys},
\emph{salts}, \emph{nonces}, \emph{initialization vectors}, \emph{challenges},
and other one-time quantities. Generating these strings can be non-trivial,
since digital system, by nature, cannot behave randomly. Compromises have
traditionally been made, using algorithms, known as pseudo-\acp{RNG}, capable
of expanding short \emph{seeds} into irregular long sequences. Unfortunately,
this has recently lead to security incidents. In fact, discovery of the
algorithm state by an attacker provides the ability to predict future `random'
quantities \cite{RFC4086}.  In some cases, \ac{RNG} attacks had a high impact,
pushing big players of information technology to incorporate true-\acp{RNG} in
their larger and newer systems \cite{taylor:spectrum-48-9}.

For simpler or legacy systems, software solutions have been developed capable
of gathering noisy data (entropy) from peripherals `perceiving' physical
phenomena complex enough to be assumed random. For instance, many operating
systems have access to mouse movements, key press timings, hard disk latencies,
etc. Such quantities are irregular, but may show correlations. Yet, they can be
accumulated in buffers from which good random numbers can be \emph{distilled}
through operations such as \emph{hashing} and the frequent \emph{re-seeding} of
pseudo-\acp{RNG} \cite{Gutterman:SP-2006}. This works fairly well for standard
computers, but may fail for little devices in an \ac{IoT} scenario. In fact,
the latter may lack most of the peripherals needed to collect entropy (for
instance, little routers, networked thermostats, and so on do not have mouses,
keyboards or hard-disks). With this, attempts at entropy distillation may even
become counterproductive, giving a false sense of security while in fact
falling back to pseudo-\ac{RNG} behavior due to lack of inputs.

The ultimate fix is clearly to bring true-\acp{RNG} even to humble devices.
Doing so in the form of hardware retrofits needs carefully weighing the
available options. Ultimately, true-\acp{RNG} are \emph{analog} systems that
harvest information from physical phenomena so intricated at the microcosmic
level to appear random at the macroscopic scale. Their simplest implementation
consists in the direct amplification and acquisition of electronic noise
(thermal, avalanche, etc). This can be effective but is inherently exposed to
tampering, since interference may be easily exchanged from genuine noise, so
that (expensive) shields can be mandatory
\cite{DomstedtElectronics:TRNG9880}. Another form of amplification is to
exploit the meta-stability of positive feedback structures. A further
alternative is to deploy phase noise in oscillators that mutually sample each
other. This lets the analog quantity be \emph{time}, so that digital voltage
levels can be adopted. Furthermore, susceptibility to tampering is improved by
the more indirect harvesting. Finally, at the highest level of sophistication,
one finds circuits based on chaotic dynamics.

The chaos-based approach is inherently different from all the others. First of
all, chaotic-\acp{RNG} replicate at the circuit level some of the features,
including \emph{sensitivity to initial conditions} \cite{Ott:CDS-1993},
conjectured to be at the root of randomness in natural systems. Secondly, they
harvest noise in a radically different way. Non-chaotic sources need to
continuously gather new noise samples, while chaotic source could in principle
use just the uncertainty in their start-up condition, thanks to sensitivity to
initial conditions.  In practice, noise will certainly be present throughout
all the operating time, but this ends up being just a bonus, since correct
functionality does not strictly demand it. For this reason, as long as it is
independent from their inner state, noise or interference collected
\emph{during} operation can often bear only minor influence on output quality
(unless it is very large, but in this case tampering can typically be detected
with ease).

\section{\acs{ADC} based true-\ac{RNG}}
Among all possible chaotic sources, discrete time ones based on the iteration
of 1-dimensional \ac{PWAM} maps \cite{Callegari:TSP-53-2} are very
appealing. The corresponding model is
\begin{equation}
  x(n+1)=M(x(n))
  \label{eq:map}
\end{equation}
where an invariant set (here normalized to $[0,1]$) exists for $M(\cdot)$.  The
\ac{PWAM} nature of $M(\cdot)$ assures that a partition $\mathcal P$ of $[0,1]$
exists such that the dynamics, once observed through a $\mathcal P$-induced
quantization function $q(x)$, can be modeled through a Markov chain
\cite{Callegari:ISCAS-2007}.

Shift type maps, such as
\begin{equation}
  M(x)=(\alpha x + \beta) \bmod 1
  \label{eq:shifts}
\end{equation}
with integer $\alpha>1$ are particularly interesting. They assure that any
subdivision of $[0,1]$ in $\alpha$ equally sized sub-intervals, wrapped around
at the invariant set boundaries, results in a Markov partition regardless of
$\beta$. Namely, the partition can be built taking any $p\in[0,1]$ and sets
$\set I_i$ for $i=0,\dots \alpha-1$, such that
\begin{equation}
  I_i=\{x:x\in[0,1] \AND i\le \alpha ((x-p) \bmod 1)< i+1\} \ .
  \label{eq:partitioninterval}
\end{equation}
The corresponding Markov chain has $\alpha$ states, with transition
probabilities among any two of them equal to $1/\alpha$. In other words, it is
identical to the chain describing the cast of an $\alpha$-faced unbiased die
\cite{Callegari:ISCAS-2007}. Thus, the system has the ability to generate
random symbols.

\begin{figure}[t]
  \begin{tightcenter}
    \begin{tabular}{cc}
      \subfloat{%
        \label{sfig:adcdac}%
        \includegraphics[scale=0.7]{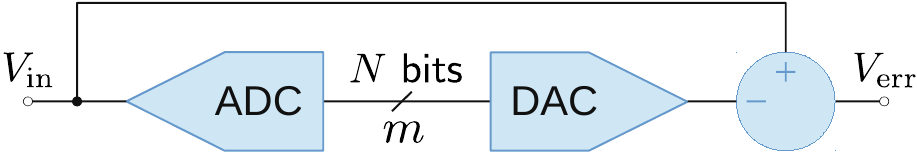}} & {\raisebox{1ex}{\small(a)}}\\
      \subfloat{%
        \label{sfig:proc}%
        \includegraphics[scale=0.7]{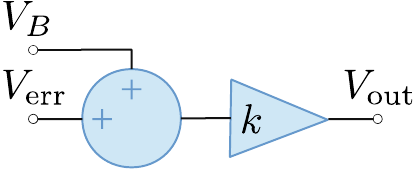}} & {\raisebox{1ex}{\small(b)}}\\
      \subfloat{%
        \label{sfig:aregister}%
        \includegraphics[scale=0.7]{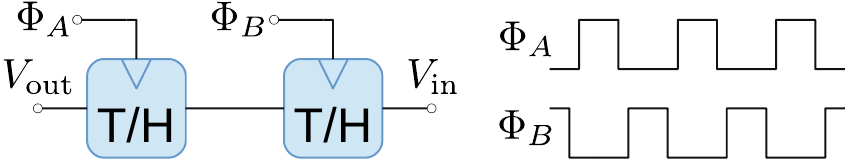}} & {\raisebox{1ex}{\small(c)}}
    \end{tabular}    
  \end{tightcenter}
  \caption{Derivation of \ac{ADC} quantization error via a cascaded \acs{DAC}
    \protect\subref{sfig:adcdac}; its processing
    \protect\subref{sfig:proc}; and the closing of the loop to create a chaotic
  with a shift-type map \protect\subref{sfig:aregister}.}
  \label{fig:ADCDAC}
\end{figure}

\begin{figure}[b]
  \begin{tightcenter}
    \includegraphics[scale=0.6]{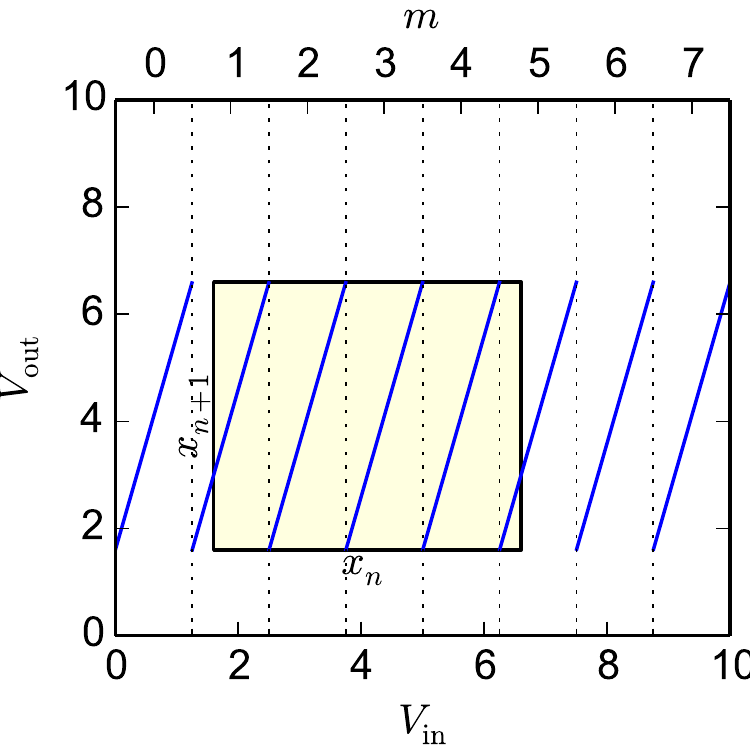}
    \caption{Sample map obtained from the quantization error of an
      \ac{ADC}. Here, $N=3$, $k=4$, $V_{\mathrm{ref}}=\unit[10]{V}$,
      $V_B=\unit[0.4]{V}$.}
    \label{fig:map}
  \end{tightcenter}
\end{figure}

Interestingly, shift maps can be easily obtained by deriving the quantization
error of an \ac{ADC}, as shown in Fig.~\subref*{sfig:adcdac}.
Let $\ADC(V)$ and $\DAC(m)$ be used to indicate the static characteristics of
the \ac{ADC} and the \ac{DAC}, respectively. For an $N$ bit single ended
\ac{ADC}, one should ideally have
\begin{equation}
  \ADC(V)=\floor{\frac{2^N}{V_{\mathrm{ref}}}V}
\end{equation}
where $V_{\mathrm{ref}}$ is the reference voltage (approximately equal to the
full scale) and a \emph{flooring} \ac{ADC} behavior is assumed. Quantization
step is $V_{\mathrm{ref}}/2^N$. For the complementary $N$ bit \ac{DAC} one has
\begin{equation}
  \DAC(m)=m \frac{V_{\mathrm{ref}}}{2^N}
\end{equation}
so that the quantization error, estimated as show in the figure is
\begin{multline}
  V_{\mathrm{err}}=V_{\mathrm{in}}- \DAC(\ADC(V_{\mathrm{in}}))=
  V_{\mathrm{in}}- \frac{V_{\mathrm{ref}}}{2^N}
  \floor{\frac{2^N}{V_{\mathrm{ref}}}V_{\mathrm{in}}}= \\
  \frac{V_{\mathrm{ref}}}{2^N} \left(\frac{2^N}{V_{\mathrm{ref}}}
    V_{\mathrm{in}} - \floor{\frac{2^N}{V_{\mathrm{ref}}} V_{\mathrm{in}}}
  \right) = \frac{V_{\mathrm{ref}}}{2^N} \left( \frac{2^N}{V_{\mathrm{ref}}}
    V_{\mathrm{in}} \bmod 1 \right) \ .
\end{multline}
The expression already contains the modulus operation that characterizes
Eqn.~\eqref{eq:shifts}. Now, assume that a further block, as shown in
Fig.~\subref*{sfig:proc} computes $V_{\mathrm{out}}=k(V_{\mathrm{err}}+V_B)$.
Altogether
\begin{equation}
  V_{\mathrm{out}}= \frac{k V_{\mathrm{ref}}}{2^N} \left( \frac{2^N}{V_{\mathrm{ref}}}
    V_{\mathrm{in}} \bmod 1 \right) + k V_B \ .
\end{equation}
Rearranging the terms, one has
\begin{multline}
  \frac{2^N}{k V_{\mathrm{ref}}} \left(V_{\mathrm{out}}-kV_B \right) =\\
  \left(k \frac{2^N}{kV_{\mathrm{ref}}} (V_{\mathrm{in}}-k V_B)+k
    \frac{2^N}{V_{\mathrm{ref}}} V_B \right) \bmod 1 \ .
  \label{eq:voltagemap}
\end{multline}
It is now sufficient to introduce the voltage-to-$x$ transformation
\begin{equation}
  x \to \frac{2^N}{k V_{\mathrm{ref}}}(V-k V_B)
  \qquad
  V \to \frac{k V_{\mathrm{ref}}}{2^N} x + k V_B
  \label{eq:transform}
\end{equation}
to see that Eqn.~\eqref{eq:voltagemap} defines the map $M(\cdot)$, with $\alpha
\equiv k$ and $\beta\equiv k 2^N V_B/V_{\mathrm{ref}}$. From this, to obtain a
dynamical system such as that in Eqn.~\eqref{eq:map}, it is enough to feed back
$V_{\mathrm{out}}$ into $V_{\mathrm{in}}$ through a delay element. The latter
can be implemented as the cascade of two \acp{TH} blocks operating in a
master-slave fashion, as in Fig.~\subref*{sfig:aregister}. With this, while $x$
spans $[0,1]$, the voltages $V_{\mathrm{in}}$ and $V_{\mathrm{out}}$ span $[k
V_B, k(V_B+V_{\mathrm{ref}}/2^N)]$. The situation is illustrated in
Fig.~\ref{fig:map}, from which it is evident that some bounds are needed for
the parameters. Specifically, $2\le k\le2^N$ (with $k\in\Nset{N}$), $V_B\ge 0$
and $k(V_B+V_{\mathrm{ref}}/2^N)\le V_{\mathrm{ref}}$, i.e., $V_B \le
V_{\mathrm{ref}}/k-V_{\mathrm{ref}}/2^N$.

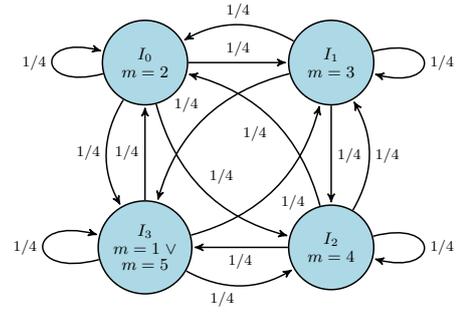
\begin{figure}
  \begin{tightcenter}
    \scalebox{0.7}{%
      \begin{tikzpicture}[->, >=stealth', shorten >=1pt, auto, node
        distance=3.5cm, thick, main node/.style={circle,fill=LightBlue, minimum
          size=1.6cm,draw}]
      
      \node[main node] (1) {\shortstack{$I_0$\\$m=2$}};
      \node[main node] (2) [right of=1] {\shortstack{$I_1$\\$m=3$}};
      \node[main node] (3) [below of=2] {\shortstack{$I_2$\\$m=4$}};
      \node[main node] (4) [left of=3] {%
        \shortstack{$I_3$\\$m=1\; \OR$\\ $m=5$}};
      
      \path[every node/.style={font=\small}]
      (1) edge [loop left] node [left] {$1/4$} (1)
      edge [] node [above] {$1/4$} (2)
      edge [bend right] node [above right] {$1/4$} (3)
      edge [bend right] node [left] {$1/4$} (4)
      (2) edge [bend right] node [above] {$1/4$} (1)
      edge [loop right] node [right] {$1/4$} (2)
      edge [] node [right] {$1/4$} (3)
      edge [bend right] node [above left] {$1/4$} (4)
      (3) edge [bend right] node [below left] {$1/4$} (1)
      edge [bend right] node [right] {$1/4$} (2)
      edge [loop right] node [right] {$1/4$} (3)
      edge [] node [below] {$1/4$} (4)
      (4) edge [] node [left] {$1/4$} (1)
      edge [bend right] node [below right] {$1/4$} (2)
      edge [bend right] node [below left] {$1/4$} (3)
      edge [loop left] node [left] {$1/4$} (4);
    \end{tikzpicture}}
  \end{tightcenter}
  \caption{Markov chain corresponding to the map in Fig.~\ref{fig:map}.}
  \label{fig:markov}
\end{figure}

Interestingly, the figure also makes evident that the \ac{ADC} quantization
inherently provides a $k$-state Markov partition for the map. In fact, the
range $[k V_B, k(V_B+V_{\mathrm{ref}}/2^N)]$ is as wide as $k
V_{\mathrm{ref}}/2^N$, and contains exactly $k$ quantization
steps. Specifically, it is sufficient to take $p=(m_{\mathrm{min}}+1)
V_{\mathrm{ref}}/2^N$ in order to have $k$ intervals such as those in
Eqn.~\eqref{eq:partitioninterval}. This is convenient, since it lets the Markov
state be directly readable from the output $m$ of the \ac{ADC}.  In detail, one
gets that the spanned $m$ values fit between
\begin{equation}
  m_{\mathrm{min}} = \floor{\frac{2^N}{V_{\mathrm{ref}}} k V_B}
 \qquad
  m_{\mathrm{max}} = \floor{\frac{2^N}{V_{\mathrm{ref}}} k V_B+k}
\end{equation}
and that system is in state $I_i$ when
\begin{equation}
  \begin{cases}
    m=m_{\mathrm{min}}+i+1 & \text{if } i<k-1\\
    m=m_{\mathrm{min}} \lor m=m_{\mathrm{max}} & \text{if } i=k-1
  \end{cases}\ .
\end{equation}
With this, one has a true-\ac{RNG} that randomly picks a value in an alphabet
of $k$ symbols at each cycle.  As an example, Fig.~\ref{fig:markov} shows the
Markov chain corresponding to the setup in Fig.~\ref{fig:map}. In view of an
efficient implementation, it is clearly convenient to make $k$ a power of $2$,
say $2^M$ with $M\in\Nset{N}$, so that each generated value fits exactly in $M$
bits. Clearly, one needs $M\le N$. In principle, $M$ should be large so that
many random bits can be generated per cycle. Yet, $M$ and $N$ must often be
contained to ease implementation.

\section{Practical implementation on a \acs{uC}}
\subsection{Implementation principles}
\begin{figure}[b]
  \begin{tightcenter}
    \begin{tabular}{cc}
      \subfloat{%
        \label{sfig:arch}%
        \includegraphics[scale=0.5]{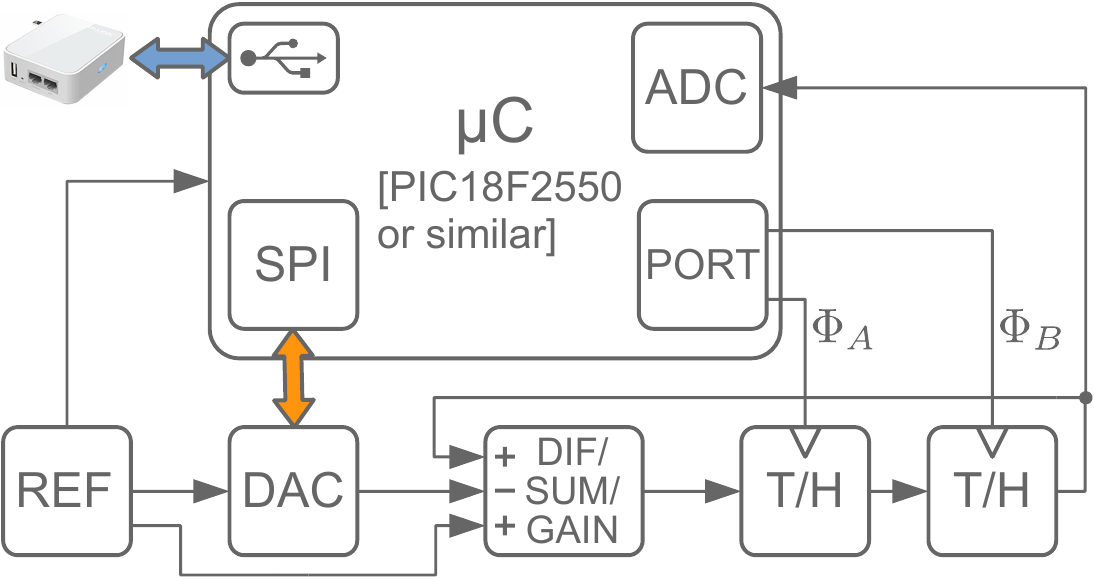}} & {\raisebox{4ex}{\small(a)}}\\
      \subfloat{%
        \label{sfig:detail}%
        \includegraphics[scale=0.5]{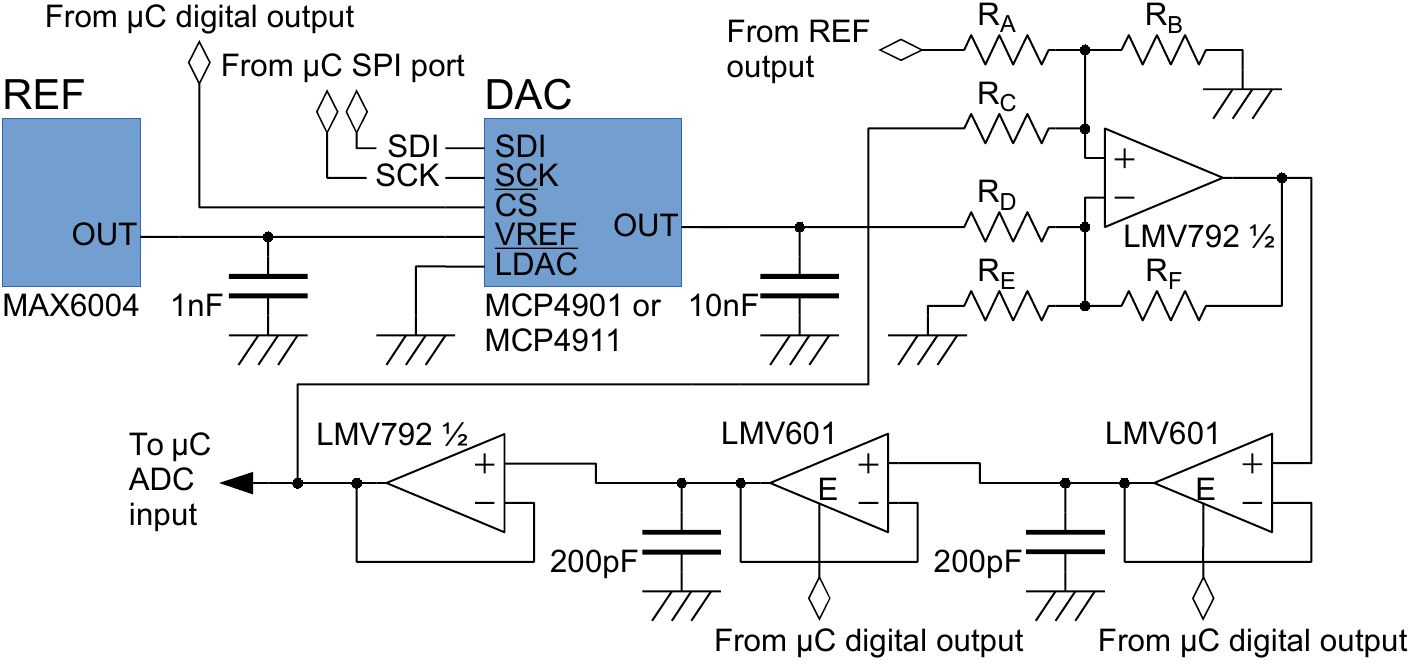}} & {\raisebox{4ex}{\small(b)}}
    \end{tabular}
  \end{tightcenter}
  \caption{\ac{uC} based architecture implementing the proposed
    true-\ac{RNG}. Block diagram \protect\subref{sfig:arch} and signal
    processing chain added to the \ac{uC} \protect\subref{sfig:detail}.}
  \label{fig:arch}
\end{figure}

Intuitively, the concepts illustrated so far can be easily materialized by a
$\ac{uC}$ based architecture as shown in Fig.~\ref{fig:arch}.
Since most \acp{uC} already include an \ac{ADC}, the number of external
components can be very low. The loop corresponding to the model in
Eqn.~\eqref{eq:map} can be closed via an external \ac{DAC} (for instance, the
MPC4901 or MPC4911, communicating with the \ac{uC} on the \acs{SPI} bus). The
loop gain and offset can be provided by a difference amplifier (for instance,
obtained by half an LVM602 or LMV792 chip), controlling $k$ and $V_B$ by means
of its resistors ($R_A$ to $R_F$ in the schematic). The loop delay requires two
\ac{TH} units. These can be obtained by \acp{OA} with an output enable feature,
followed by suitably large capacitors (for instance each \ac{TH} can be
obtained by an LVM601). A voltage reference is required for accurate
operation. Some \acp{uC} readily provide one, otherwise a dedicated chip (such
as the MAX6004) can be used. In the proposed architecture, the loop path goes
through the \ac{uC}, which acquires the \ac{TH} output and passes the
corresponding numerical data to the external \ac{DAC} via its \acs{SPI}
bus. The \ac{uC} is also in charge of generating the timing signals for the
\ac{DAC} and the \ac{TH} units. These are transferred via 3 lines on one of the
digital I/O ports.  For prototyping a stand alone unit, the PIC18F2550 \ac{uC}
can conveniently be selected, in view of its low price and integrated \acs{USB}
bus, which favors interfacing. In fact, many little network devices include
\acs{USB} connection facilities. Alternatively, the portion of schematic in
Fig.~\subref*{sfig:detail} can be added to an existing design including a
\ac{uC} with some spare computational resources. With this, costs can be
further lowered.

A notable advantage of the proposal is that it makes $m$ readily available
inside the \ac{uC}. Consequently, at each cycle, the \ac{uC} can compute a
random symbol from it while passing it to the \ac{DAC}. The so generated
symbols can be progressively accumulated and packed in bytes. When the \ac{uC}
has buffered enough of them (or when a request arrives), they can be
transferred to the embedding system via one of the \ac{uC} communication
interfaces, for instance, the \ac{USB} bus, possibly under encryption to
prevent snooping or tampering.

\subsection{Implementation requirements}
In practice, the implementation cannot be so direct and requires some
care. Specifically, it is not possible to straightforwardly deploy the block
diagram in Fig.~\ref{fig:ADCDAC}, because of implementation uncertainties. For
instance, in most \ac{uC}, the \ac{ADC} \acp{lsb} can be quite erratic due to
couplings with digital lines. Furthermore, accuracy in the setting of $V_B$
should be in the order of half a quantization step, but this is too fine to
achieve when the \ac{ADC} resolution is 8--10 bits.  Any implementation attempt
ignoring these matters would most certainly fail. Fortunately, the two issues
can be easily worked around by artificially degrading the \ac{ADC} resolution
before computing $V_\mathrm{err}$. This can be done as follows. Assume that
$\hat N$ is the nominal \ac{ADC} resolution and that $\hat m$ is the actual
$\hat N$-bit \ac{ADC} output at each cycle, then let $N$ be just \emph{a
  portion} of $\hat N$, so that $m$ can be obtained by shifting $\hat m$ right
by $\hat N-N$ bits. With this, the value to be passed to the \ac{DAC} can be
obtained by back shifting $m$ to the left with zero padding. Incidentally, this
permits the use of an external \acp{DAC} with a lower resolution $\tilde N<\hat
N$, as long as $\tilde N\ge N$. Good $N$ values can be $3$ to $5$
bits. Furthermore, $M$ (which determines $k=2^M$) must be taken to be strictly
less than $N$. In fact, this lets $V_B$ be chosen so that the
$V_{\mathrm{out}}$ range lies \emph{with clearance} within the \ac{ADC} input
range and the \ac{OA} output saturation voltages. Good $M$ values are $2$ to
$4$ bits, depending on $N$, while good $V_B$ values let $k V_B$ fall halfway
between voltages corresponding to transition points for $m$.

\begin{figure}
  \begin{tightcenter}
      \subfloat[\label{sfig:experiment-map}]{%
        \includegraphics[scale=0.45]{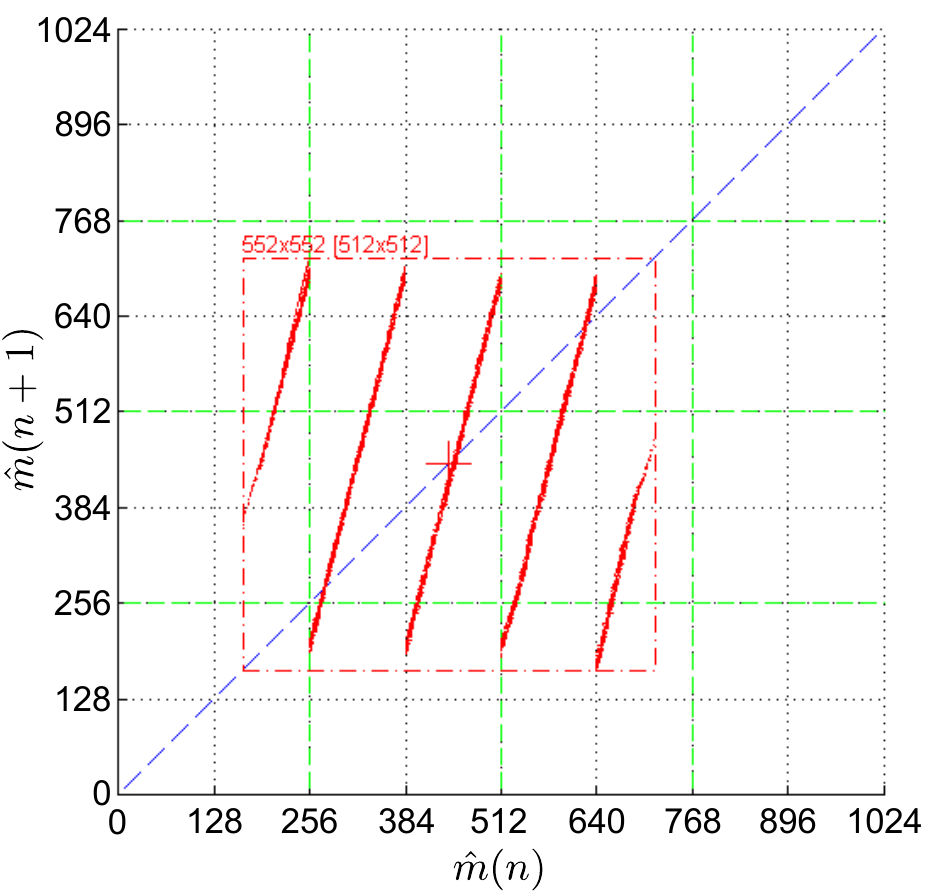}}\quad 
      \subfloat[\label{sfig:probs}]{%
        \includegraphics[scale=0.45]{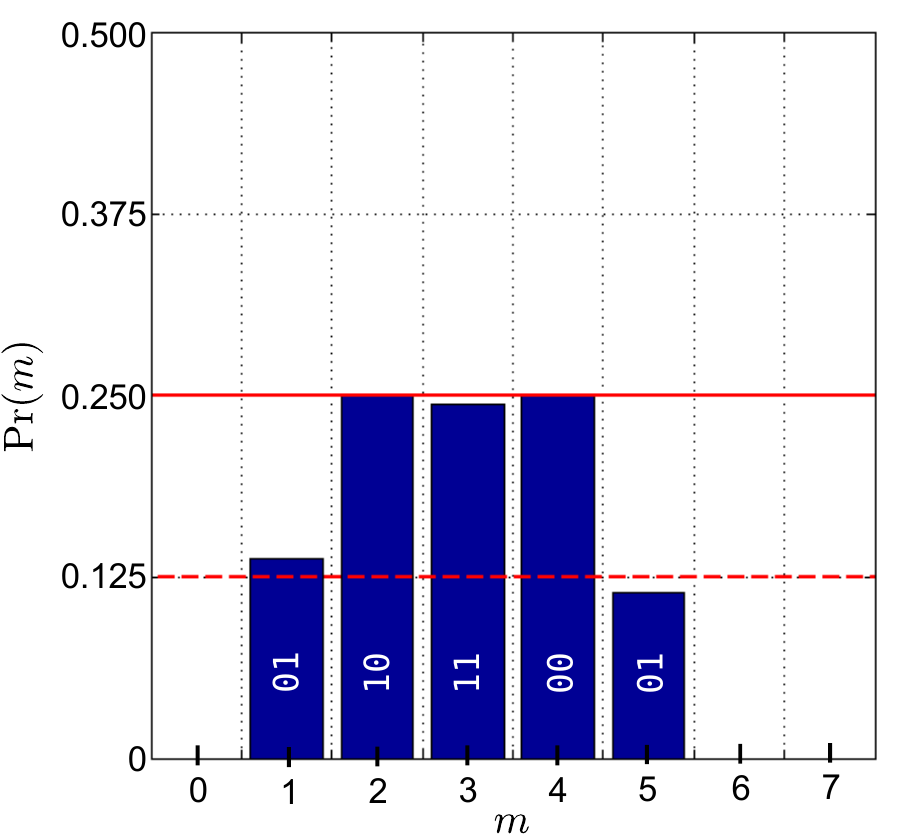}}
  \end{tightcenter}
  \caption{Experimental evaluation of the system prototype illustrated in
    Fig.~\ref{fig:demo}, set up for $\hat N=10$, $N=3$, $M=2$,
    $V_{\mathrm{ref}}=\unit[4.096]{V}$, $V_B=\unit[192]{mV}$. Left: shape of the
    implemented $M(x)$, reconstructed by plotting $\hat m(n+1)$ vs $\hat
    m(n)$. Right: statistical distribution of $m$.}
  \label{fig:experiment}
\end{figure}

An advantageous side effect of the artificial resolution degradation is that it
lets the \ac{uC} observe the chaotic system state $x$ via $\hat m$, i.e., at a
resolution $\hat N$ higher than the $N$ bits required for operation. In other
words, one gains insight on the \emph{analog} system state without the need for
external instrumentation. Such an opportunity can be useful in view of self
diagnostics or automatic detection of tampering attempts. As an example,
Fig.~\subref*{sfig:experiment-map} shows a reconstruction of the iterated map
obtained by \emph{actual experimental data} acquired by the \ac{uC} itself.

\subsection{Experimental validation}
The proposed architecture has been extensively tested, using the prototype
illustrated in Fig.~\ref{fig:demo}. For the tests, parameters have been set as
in the caption of Fig.~\ref{fig:experiment}.  Pane~\subref{sfig:probs} shows
the experimentally obtained $m$ values and their relative frequencies. The
numbers on the graph columns indicate the output bits associated to each $m$
value (obtained as $m \bmod 4$). Their occurrence probability is well balanced
at approximately $1/4$. Long output streams have been tested for their
information content, achieving $>0.99$ bits of entropy for each delivered
output bit (estimated via a marginal entropy approach). The NIST 800-22 suite
\cite{NIST:SP800-22} has also been applied. The corresponding tables of
outcomes cannot fit in this paper, but are worth summarizing. The \emph{raw}
generator is incapable of passing \emph{all} tests (though it can already pass
many of them). Yet, a minimal post-processing (by a Von Neumann corrector or
the XORing of consecutive bits) is already sufficient for full
conformance. Similarly, when the raw data produced by the generator is used to
feed the entropy distiller in the Linux \ac{OS}, the distiller output can pass
all tests.

\section{Conclusions}
\label{sec:conclusions}
A new architecture of a \ac{uC} based true-\ac{RNG} has been proposed,
prototyped and tested. Internally it relies on chaotic dynamics. Expected cost
is extremely low and many interconnection options are possible, making it
suitable as a retrofit for existing networked units that lack a true-\ac{RNG}
and that may be exposed to \ac{RNG} attacks.

\bibliographystyle{SC-IEEEtran}
\bibliography{macros,IEEEabrv,config,various,sensors,analog,chaos,ultrasound}

\begin{thebibliography}{10}
\providecommand{\doi}[1]{DOI:#1}
\providecommand{\url}[1]{#1}
\csname url@samestyle\endcsname
\providecommand{\newblock}{\relax}
\providecommand{\bibinfo}[2]{#2}
\providecommand{\BIBentrySTDinterwordspacing}{\spaceskip=0pt\relax}
\providecommand{\BIBentryALTinterwordstretchfactor}{4}
\providecommand{\BIBentryALTinterwordspacing}{\spaceskip=\fontdimen2\font plus
\BIBentryALTinterwordstretchfactor\fontdimen3\font minus
  \fontdimen4\font\relax}
\providecommand{\BIBforeignlanguage}[2]{{%
\expandafter\ifx\csname l@#1\endcsname\relax
\typeout{** WARNING: IEEEtran.bst: No hyphenation pattern has been}%
\typeout{** loaded for the language `#1'. Using the pattern for}%
\typeout{** the default language instead.}%
\else
\language=\csname l@#1\endcsname
\fi
#2}}
\providecommand{\BIBdecl}{\relax}
\BIBdecl

\bibitem{Gartner:IOT-2013}
P.~Middleton, P.~Kjeldsen, and J.~Tully, ``Forecast: The internet of things,
  worldwide, 2013,'' Gartner, Report G00259115, Nov. 2013.

\bibitem{roman:computer-44-9}
R.~Roman, P.~Najera, and J.~Lopez, ``Securing the internet of things,''
  \emph{{IEEE} Computer}, vol.~44, no.~9, p. 51–58, 2011.

\bibitem{Lynn:PCMAG-2011-05}
\BIBentryALTinterwordspacing
S.~Lynn, ``Survey: Biz network devices vulnerable, almost obsolete,'' \emph{PC
  Magazine (Online)}, May 2011. [Online]. Available:
  \url{http://www.pcmag.com/article2/0,2817,2385833,00.asp}
\BIBentrySTDinterwordspacing

\bibitem{Geer:SOT-2014}
\BIBentryALTinterwordspacing
D.~Geer, ``Security of things,'' presented at the Security of Things Forum,
  Cambridge, MA, USA, May 2014, keynote speech. [Online]. Available:
  \url{http://geer.tinho.net/geer.secot.7v14.txt}
\BIBentrySTDinterwordspacing

\bibitem{wikipedia:RNGattack}
\BIBentryALTinterwordspacing
Wikipedia, ``Random number generator attack,'' 2014, accessed 2014-05-13.
  [Online]. Available:
  \url{http://en.wikipedia.org/wiki/Random_number_generator_attack}
\BIBentrySTDinterwordspacing

\bibitem{taylor:spectrum-48-9}
G.~Taylor and G.~Cox, ``Digital randomness,'' \emph{{IEEE} Spectr.}, vol.~48,
  no.~9, p. 32–58, 2011.

\bibitem{DomstedtElectronics:TRNG9880}
\BIBentryALTinterwordspacing
B.~D. Electronics, ``{TRNG9880} random number processing,'' Dec. 2013, white
  paper. [Online]. Available:
  \url{www.trng98.se/getfile.php?file=trng9880_info.pdf}
\BIBentrySTDinterwordspacing

\bibitem{Pareschi:ISCAS09}
F.~Pareschi, G.~Scotti, L.~Giancane, R.~Rovatti, G.~Setti, and A.~Trifiletti,
  ``Power analysis of a chaos-based random number generator for cryptographic
  security,'' in \emph{Proc. IEEE Intern. Symp. on Circuits and Systems, 2009},
  2009, pp. 2858--2861.

\bibitem{Callegari:IJCTA-33-1}
S.~Callegari, R.~Rovatti, and G.~Setti, ``First direct implementation of true
  random source on programmable hardware,'' \emph{International Journal of
  Circuit Theory and Applications}, John Wiley \& Sons, vol.~33, no.~1, pp.
  1--16, 2005.  \doi{10.1002/cta.301}

\bibitem{Callegari:TSP-53-2}
S.~Callegari, R.~Rovatti, and G.~Setti, ``Embeddable {ADC}-based true random
  number generator for cryptographic applications exploiting nonlinear signal
  processing and chaos,'' \emph{{IEEE} Trans. Signal Process.}, vol.~53, no.~2,
  pp. 793--805, Feb. 2005.  \doi{10.1109/TSP.2004.839924}

\bibitem{Callegari:ISCAS-2007}
S.~Callegari and G.~Setti, ``{ADCs}, chaos and {TRNGs}: a generalized view
  exploiting {M}arkov chain lumpability properties,'' in \emph{Proc.\@ of
  ISCAS}, New Orleans, LA (USA), May 2007, pp. 213--216.
  \doi{10.1109/ISCAS.2007.378314}

\bibitem{Callegari:ISCAS-2008}
S.~Callegari, ``Introducing \emph{Complex Oscillation Based Test}: an
  application example targeting analog to digital converters,'' in
  \emph{Proc.\@ of ISCAS}, Seattle, WA (USA), May 2008, pp. 320--323.
  \doi{10.1109/ISCAS.2008.4541419}

\bibitem{RFC4086}
\BIBentryALTinterwordspacing
D.~E. Eastlake, J.~I. Shiller, and S.~D. Crocker, ``Randomness requirements for
  security,'' RFC 4086, Internet Engineering Task Force, 2005. [Online].
  Available: \url{http://www.ietf.org/rfc/rfc4086.txt}
\BIBentrySTDinterwordspacing

\bibitem{Gutterman:SP-2006}
Z.~Gutterman, B.~Pinkas, and T.~Reinman, ``Analysis of the linux random number
  generator,'' in \emph{Proc. of the IEEE Symposium on Security and Privacy},
  Oakland, CA, USA, May 2006.

\bibitem{Ott:CDS-1993}
E.~Ott, \emph{Chaos in dynamical systems}.\hskip 1em plus 0.5em minus
  0.4em\relax Cambridge University Press, 1993.

\bibitem{NIST:SP800-22}
\BIBentryALTinterwordspacing
A.~Rukhin, J.~Soto, J.~Nechvatal, M.~Smid, E.~Barker, S.~Leigh, M.~Levenson,
  M.~Vangel, D.~Banks, A.~Heckert, J.~Dray, and S.~Vo, \emph{A Statistical Test
  Suite for Random and Pseudorandom Number Generators for Cryptographic
  Applications}, National Institute for Standards and Technology Special
  publication 800-22, May 2001. [Online]. Available:
  \url{http://csrc.nist.gov/rnd/SP800-22b.pdf}
\BIBentrySTDinterwordspacing

\end{thebibliography}

\end{document}